\documentclass[1p]{elsarticle}
\graphicspath{{Pictures/}} 
\usepackage[export]{adjustbox}
\usepackage{graphicx}
\graphicspath{ {images/} }
\usepackage{amssymb}
\usepackage{bold-extra}
\usepackage{amsmath} 
\usepackage{color}
\usepackage{slashed}
\usepackage{adjustbox,lipsum}
\usepackage{array}
\usepackage{geometry}
\usepackage{bm}
\usepackage{commath}

\begin{document}

\begin{frontmatter}
\title{Beam polarization effects on top-pair production at the ILC}
\author{Nhi M. U. Quach$^{a,b *}$, Yoshimasa Kurihara$^{b}$, 
	\vspace{0mm}Khiem H. Phan$^{c}$, Takahiro Ueda $^{d}$}
\address{$^a$ The Graduate University for Advanced Studies (SOKENDAI), Hayama, Kanagawa 240-0193, Japan. }
\address{$^b$ High Energy Accelerator Research Organization (KEK), Tsukuba, Ibaraki 305-0801, Japan. }
\address{$^c$University of Science Ho Chi Minh City, 227 Nguyen Van Cu, Dist.5, Ho Chi Minh City, Vietnam.}
\address{$^d$Nikhef, Science Park 105 1098 XG Amsterdam, Netherlands.}

\begin{abstract}
Full one-loop electroweak-corrections for an $e^-e^+\rightarrow t \bar{t}$ process  associated with sequential $t\rightarrow b \mu\nu_\mu$ decay are discussed.
At the one-loop level, the spin-polarization effects of the initial electron and positron beams are included in the total and differential cross sections.
A narrow-width approximation is used to treat the top-quark production and decay while including full spin correlations between them.
We observed that the radiative corrections due to the weak interaction have a large polarization dependence on both the total and differential cross sections.
Therefore, experimental observables that depend on angular distributions such as the forward-backward asymmetry of the top production angle must be treated carefully including radiative corrections. 
We also observed that the energy distribution of bottom quarks is majorly affected by the radiative corrections.
\end{abstract}
\end{frontmatter}

%
%
\section{Introduction}
The discovery of the Higgs boson\cite{Aad:2012tfa,Chatrchyan201230} in 2012 showed the standard theory of particle physics to be well established.
Even though the standard theory can describe the microscopic nature at a subatomic level very precisely\cite{Olive:2016xmw}, it cannot be the most fundamental theory of nature because, for instance, it includes many parameters (e.g., particle masses and couplings, number of generations) that are not determined within the theory.
While experiments at the Large Hadron Collider continue to search for signals beyond the standard model (BSM), non have been reported to date\footnote{For the most up-to-date results, see \cite{ATLASweb} for ATLAS and \cite{CMSweb} for CMS collaborations.}.
Besides discovering new particles, pursuing the BSM also involves precise measurements of the properties of known particles.
Milestones along this direction must surely be the Higgs boson and the top quark.
Because the top quark is the heaviest fermion with a mass above even the electroweak symmetry-breaking scale, it is naturally expected to play a special role in the BSM\@.
In addition, it has been pointed out that the vacuum stability of the Higgs potential depends strongly  on the Higgs and top-quark masses\cite{Alekhin2012214}. 
Hence, the precise measurement of top-quark properties is crucial for understanding the stability of the universe, as well as in the search for BSM signals.
		
The International Linear Collider(ILC)\cite{Behnke:2013xla}, which is a proposed electron positron colliding experiment with centre-of-mass (CM) energies above 250 GeV, is being discussed intensively as a future project in high-energy physics. 
The main goals of ILC experiments would be a precise measurement of the Higgs and top-quark properties as well as searching directly for new particles.
The ILC will use spin-polarized beams for both its electron and positron beams\cite{Adolphsen:2013jya,Adolphsen:2013kya}  to increase its sensitivity to new physics and to improve its measurement accuracy.
For instance, for many processes, beam polarization is a simple way to increase the signal cross section while suppressing the background.
Moreover, beam polarization allows new properties to be measure (e.g., the polarization dependence of cross sections). 
Detailed Monte Carlo studies have shown that the ILC would be able to measure most of the standard model parameters to within sub-percent levels\cite{Baer:2013cma}.

Because of the improved experimental accuracy intended of the ILC, theoretical predictions must be given with new level of precision.
In particular, a radiative correction due to the electroweak interaction (including spin polarizations) is mandatory for such requirements.
Before the discovery of the top quark, a full electroweak radiative correction was conducted for an $e^-e^+\rightarrow t\bar{t}$ process at a lower energy\cite{Fujimoto:1987hu}, and was then obtained independently for higher energies\cite{Fleischer:2002rn,Fleischer2003}.
The same correction including radiative photon, $e^-e^+\rightarrow t\bar{t}\gamma$ process, has also been reported\cite{Khiem2013}.
However, none of previous calculations include the effect of spin polarization.
In the present study, we report full electroweak radiative corrections for the process 
$e^-e^+\rightarrow t \bar{t}\rightarrow b \bar{b} \mu^+\mu^-\nu_{\mu} \bar{\nu}_{\mu}$
using a narrow-width approximation for the  top quarks.
Spin-polarization effects are included, not only in the initial beams  but also in the full spin correlations of the production and decay of top quarks.

This report is organized as follows. 
The calculation method is explained in Section \ref{CM}. 
We use the GRACE-Loop system to calculate cross sections.
A system-checking method is also explained in Section \ref{CM}.
In Section \ref{R&D}, we show results of electroweak corrections of the total cross section as well as the angular distribution with spin-polarized beams. 
The effects of radiative corrections on top-quark decay products, including a spin correlation, are also discussed using a narrow-width approximation. 
The contribution of an NLO-QCD correction is briefly discussed  in Section \ref{R&D}. 
We summarize and conclude this report in Section \ref{summary}. 
In \ref{ap}, we summarize the formulae of the NLO-QCD correction for massive quark production.

%
%
\section{Calculation Method}\label{CM}
	\begin{table}[b]
		\begin{center}
			\begin{tabular}{|c||c|c|}
				\hline 
				$$ & $(C_{UV},{\lambda},{\tilde{\alpha},\tilde{\beta},\tilde{\delta},\tilde{\kappa},\tilde{\epsilon}})$ & $(C_{UV},{\lambda},{\tilde{\alpha},\tilde{\beta},\tilde{\delta},\tilde{\kappa},\tilde{\epsilon}})$\\
				$$ & $(0,10^{-17},0,0,0,0,0)$ & $(0,10^{-17},10,20,30,40,50)$\\
				\hline
				$e^-_L e^+_R$ 
&$-0.\underline{2471410165636298979812945}17$ & $-0.\underline{2471410165636298979812945}91$\\
				$e^-_R e^+_L$ 
&$-0.\underline{1039819068011748338895468}68$ & $-0.\underline{1039819068011748338895468}51$\\
				\hline
			\end{tabular}
			\caption{Non linear gauge-parameter independence of amplitude. 
The results are stable over 25 digits using quartic-precision variables.}\label{CUVindependence}
		\end{center}
	\end{table}
For precise cross-section calculations of the target process in this study,  we used the GRACE-Loop system, which is an automatic system for calculating cross sections of scattering processes at one-loop level for the standard theory\cite{Belanger2006117} and the minimal supersymmetric standard model\cite{PhysRevD.75.113002}. 
This system has been used to treat  electroweak processes with two, three, or four particles in the final state\cite{Belanger2003152,Belanger2003163, Belanger2003353,Kato:2005iw}.
The GRACE-Loop system has the following features.
Firstly, the renormalization of the electroweak interaction is carried out using an on-shell scheme\cite{doi:10.1143/PTPS.73.1,doi:10.1143/PTPS.100.1}. 
Secondly, the infrared divergences are regulated using a fictitious photon mass $\lambda$\cite{doi:10.1143/PTPS.100.1}.
Thiedly, the symbolic manipulation system FORM\cite{Vermaseren:2000nd} is used to handle all Dirac and tensor algebras in $n$ space time dimensions. 
Fourthly, GRACE generates FORTRAN source code that calls library subroutines to calculate the scattering amplitudes.
Fifthly, for loop integrations, all tensor one-loop integrals are reduced to scalar integrals using our own formalism, whereupon the integrations are performed  using packages FF\cite{vanOldenborgh:1990yc} and LoopTools\cite{Hahn:1998yk}. 
Finally, phase-space integrations are done using an adaptive Monte Carlo integration package BASES\cite{Kawabata1986127,KAWABATA1995309}.
For numerical calculations, we use quartic precision for floating-point variables. 

To treat spin polarization in loop calculations, we apply the projection operators on fermion wave functions. 
A spin projection of the initial beams is realized simply by multiplying the spin-projection operator$P_{\lambda}=\frac{1}{2}(1+\lambda\gamma_5\slashed{p}/m)$, where $p$ is the four-momentum of beam particles and $\lambda=\pm 1$ is their helicity.
Here, we assume that initial beams comprise light fermions with no transverse momenta.
The electron/positron completeness relation becomes $\sum_s u(p)^s\bar{u}^s(p)=\frac{1}{2}(1+\lambda\gamma_5)(\slashed{p}+m)$.
For top quarks, the spin polarization vector can be taken as 
\begin{eqnarray*}
s^{\mu}_t&=&\left(
\frac{\bm{p}_t\cdot\hat{\bm{s}}_t}{m_t}, \hat{\bm{s}}_t+ \frac{ (\bm{p}_t\cdot\hat{\bm{s}}_t)\bm{p}_t }{m_t(E_t+m_t)}
\right),
\end{eqnarray*}
where $m_t$ is the top-quark mass, $E_t$ is the top-quark energy, and $\bm{p}_t$ is the top-quark three-momentum.
The spin id projected  in the direction of the top-quark momentum direction using a direction vector 
$\hat{\bm{s}}_t=\bm{p}_t/|\bm{p}_t|$.
The completeness relations in this case are given as $\sum_\lambda u(p,\lambda)\bar{u}(p, \lambda) =\frac{1}{2}(1+\lambda\gamma_5\slashed{s})(\slashed{p}+m)$ for top quarks and $\sum_\lambda v(p,\lambda)\bar{v}(p, \lambda) =\frac{1}{2}(1+\lambda\gamma_5\slashed{s})(\slashed{p}-m)$ for anti-top quarks.

In GRACE, while using the $R_\xi$-gauge in the linear gauge-fixing terms, the non linear gauge-fixing Lagrangian\cite{Boudjema:1995cb,Belanger2006117} is employed, namely
\begin{eqnarray*}
\mathcal{L}_{GF}&=&
-\frac{1}{\xi_W}\left|\left(
\partial_{\mu}-ie {\bm{\tilde{\alpha}}} A_{\mu}-igc_W\bm{\tilde{\beta}} Z_{\mu}\right)W^{\mu+}
+\xi_W \frac{g}{2}\left(v+\bm{\tilde{\delta}} H + i\bm{\tilde{\kappa}}\chi_3\right)\chi^+\right|^2\\
&~&-\frac{1}{2\xi_Z}\left(\partial\cdot Z + \xi_Z\frac{g}{2c_W}
\left(v+\bm{\tilde{\varepsilon}} H\right)\chi_3\right)^2-\frac{1}{2\xi_A}\left(\partial\cdot A\right)^2,
\end{eqnarray*}
for the sake of system checking.
Here $A,Z,W,\chi$, and $H$ denote the wave functions of the corresponding fields, and $\xi$\rq{}s are gauge parameters for the linear gauge-fixing terms.
The results must be independent of the non linear gauge parameters $\{\bm{\tilde{\alpha},\tilde{\beta},\tilde{\delta},\tilde{\kappa},\tilde{\varepsilon}}\}$. 
We can perform system checking to confirm the correctness of the system. 
Before calculating cross sections, we checked for ultra-violet coefficient ($C_{UV}$) independence, photon-mass ($\lambda$) independence, and gauge invariance numerically at several randomly-chosen phase points. 
for instance, in the polarized case at a CM energy of 500 GeV, we confirmed ultra-violet coefficient and photon-mass independence, both with stable results over 19 digits, when parameters $C_{UV}$ and $\lambda$ changed by three orders of magnitude from their nominal values.
Meanwhile, the non linear gauge-invariance results are stable over 25 digits against changing those values, as shown in Table \ref{CUVindependence}.
We note that the parameter dependence of the amplitude is logarithmic for $C_{UV}$ and $\lambda$, wheres it is up to quartic for the non linear gauge parameters.

In addition to the above checks, we examined the soft-photon cut-off independence: for cross sections at the one-loop level,the  results must be independent of a hard-photon cut-off parameter $k_c$. 
We confirmed that the integration results are self-consistent within the statistical error of numerical phase-space integrations while varying $k_c$ from $10^{-4}$ GeV to $10^{-1}$ GeV.  

%
%
\section{Results and discussions}\label{R&D}
For cross-section calculations of the production process $e^-e^+ \rightarrow t \bar{t}$ and its sequential top decay, we use the input parameters listed in Table \ref{parameters}. 
The masses of the light quarks (i.e., other than the top quark) and $W$ boson are chosen to be consistent with low-energy experiments\cite{Khiem2015192}.
Other particle masses are taken from recent measurements\cite{Olive:2016xmw}.
The weak mixing angle is obtained using the on-shell condition $\sin^2{\theta_W} = 1- m_W^2/m_Z^2$ because of our renormalization scheme.
The fine-structure constant $\alpha=1/137.0359859$ is taken from the low-energy limit of Thomson scattering, again because of our renormalization scheme.
The $W$-boson width is taken as a calculated value at tree level using the same parameters as above.
	\begin{table}[b]
		\begin{center}
			\begin{tabular}{|c|c||c|c|}
\hline
$u$-quark mass & $58.0\times10^{-3}$ GeV & $d$-quark mass & $58.0\times10^{-3}$ GeV \\
$c$-quark mass & $1.5$ GeV & $s$-quark mass & $92.0\times10^{-3}$ GeV \\
$t$-quark mass & $173.5$ GeV & $b$-quark mass & $4.7$ GeV \\
$Z$-boson mass & $91.187$ GeV & $W$-boson mass & $80.370$ GeV \\
Higgs mass & 126 GeV & $W$-boson width & $1.993$ GeV \\
\hline
			\end{tabular}
			\caption{Input parameters}
\label{parameters}
		\end{center}
	\end{table}
\subsection{Production Cross Sections}
We focus on CM energies from $500$-$1000$ GeV to avoid possible complications from large QCD corrections near the production threshold. 
In the energy region, a target of top-quark physics is a precise measurement of the $Z$-top and top-Yukawa couplings.
It is reasonable to  expect that information beyond the standard theory could be probed through precise measurements of a top-production form factor\cite{Khiem:2015ofa}.
To extract new physics from the form-factor measurement, one has to understand precisely the effects of higher-order corrections on the measurements.

\begin{figure}[t]
  	\begin{center}
  		\includegraphics[width={10cm}]{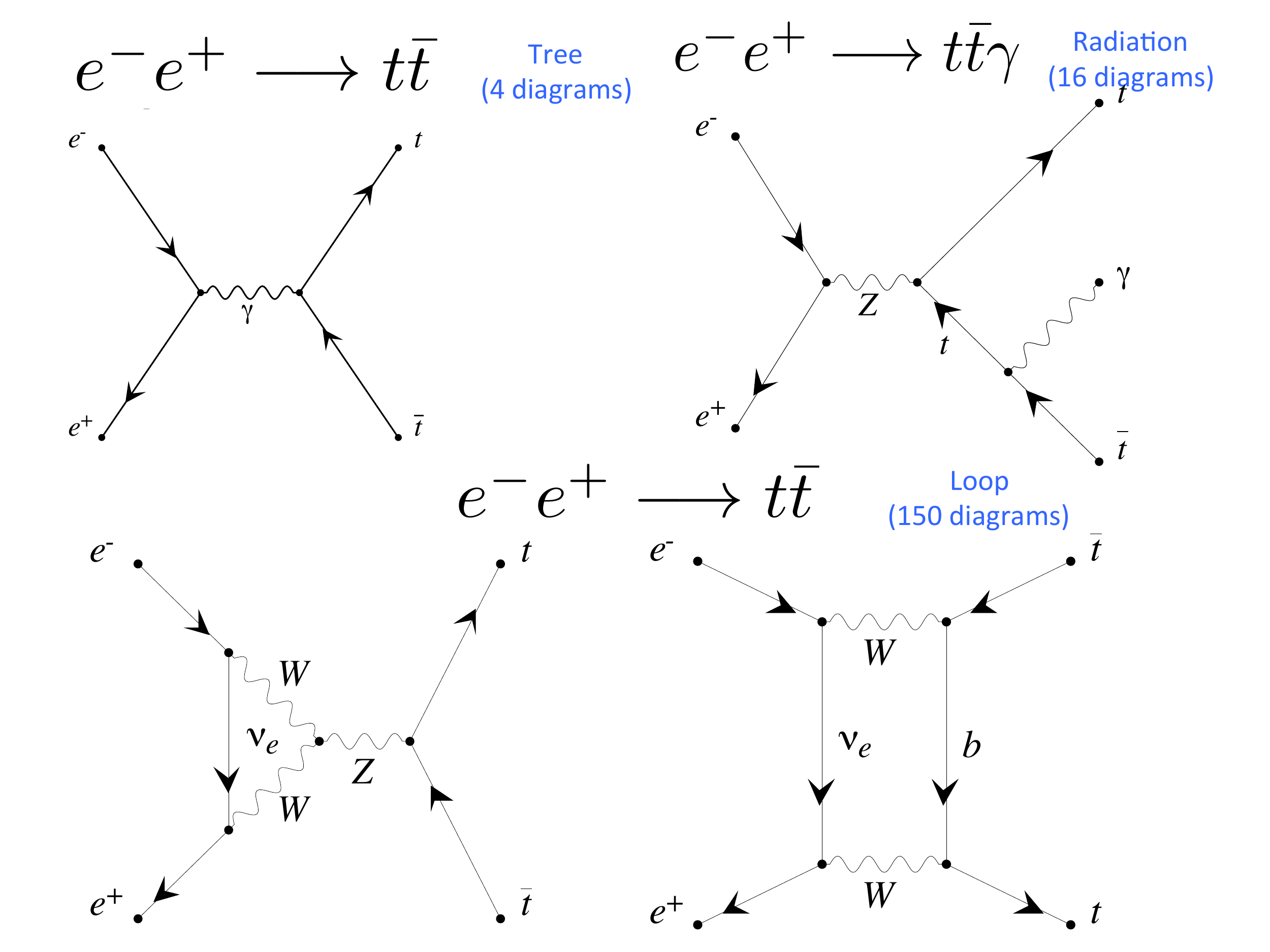}
  	\end{center}
  	\caption{Examples of Feynman diagrams for $e^-e^+ \rightarrow t \bar{t}$ at tree level, with real radiation and at loop level.}
  	\label{Feynman diagrams}
  \end{figure}

\begin{figure}[t]
  	\begin{center}
  		\includegraphics[width={\linewidth}]{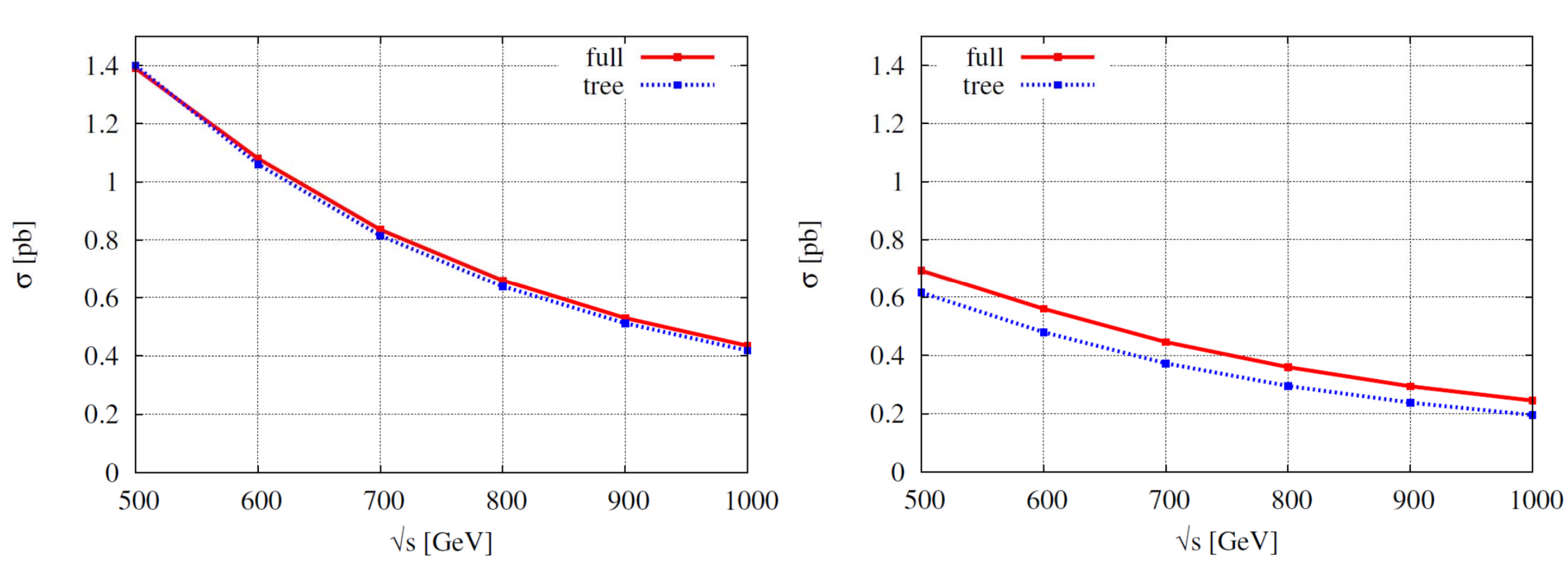}
		\includegraphics[width={\linewidth}]{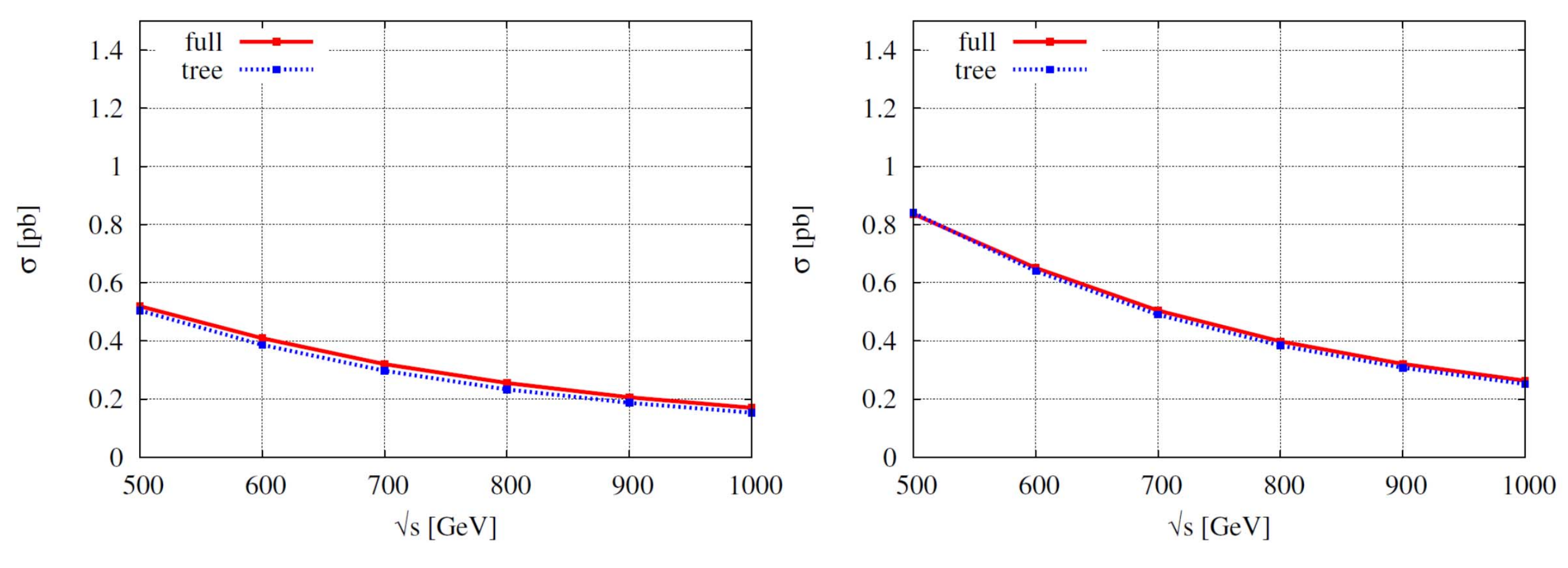}
  	\end{center}
  	\caption{Total cross sections with respect to the CM energy $\sqrt{s}$ from $500$ GeV to $1000$ GeV, assuming $100\%$ of $e^-_L$ and $e^+_R$ for the upper-left figure, and vice versa ($e^-_R$ and $e^+_L$) for the upper-right figure.
Lower-left and lower-right figures show cross sections with non-polarization and polarization with a design value ($e^-_L=80\%$ and $e^+_R=30\%$, respectively).
The dotted lines show the results for the tree level while the solid lines correspond to the full one-loop electroweak correction.}
  	\label{fig:4.2}
  	\end{figure}

\begin{figure}[t]
  	\begin{center}
  		\includegraphics[width={12cm}]{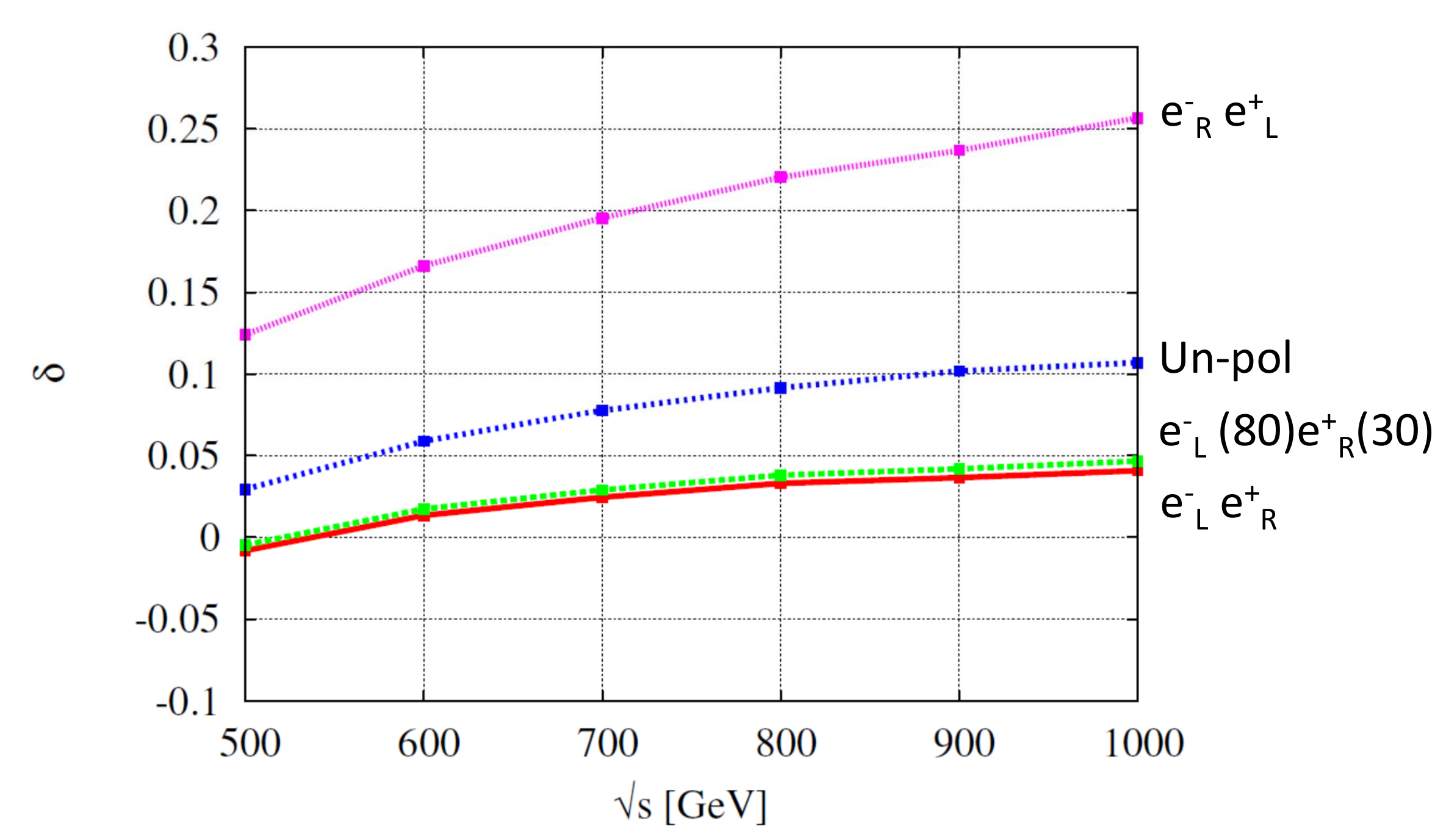}
  		\caption{Ratio of the full correction $\delta$ for various polarization conditions. 
From the top of the figure, the lines show $e^-_R e^+_L$ polarization, non-polarization, design polarization, and $e^-_L e^+_R$ polarization, in that order.}
  		\label{deltaNLO}
  	\end{center}
  \end{figure}

For the $e^-e^+ \rightarrow t \bar{t}$ process, there are four Feynman diagrams at tree level, $16$ with real-photon radiation, and $150$ at the one-loop level. 
Typical diagrams are shown in  Fig~\ref{Feynman diagrams}.
We calculate the total cross sections as a function of CM energy of $500$-$1000$ GeV assuming $100\%$ left-hand polarization for electrons ($e^-_L$) and $100\%$ right hand polarization positrons ($e^+_R$), or vice versa ($e^-_R$ and $e^+_L$).
We omit cross sections involving $e^-_L e^+_L$ and $e^-_R e^+_R$ collisions because they yield negligible contributions. 
The cross sections so obtained are shown in Fig.~\ref{fig:4.2} as functions of the colliding energy.
As shown in the upper panels of Fig.~\ref{fig:4.2}, the total cross sections for the $e^-_Le^+_R$ collision are roughly twice those for the $e^-_Re^+_L$ collision due to the $P$-violation of the weak interaction.
When design values of polarizations ($e^-_L=80\%$ and $e^+_R=30\%$) can be realized at the ILC, we will gain roughly $50\%$ in total cross section compared with the non-polarized case.
In addition, the total amount of electroweak corrections is smaller for the $e^-_Re^+_L$ case than that for the $e^-_Le^+_R$ case.
For a simple evaluation of the fraction of higher-order corrections, let us introduce a ratio $\delta=(\sigma_{NLO}-\sigma_{Tree})/\sigma_{Tree}$, where $\sigma_{NLO}$ and $\sigma_{Tree}$ are the total cross sections at a full $\cal{O}(\alpha)$ correction and that at tree level, respectively.  
The results so obtained are summarized in Fig.~\ref{deltaNLO}.
For instance, at a CM energy of $500$ GeV, the electroweak correction of $e^-_L e^+_R$  is $-0.8\%$ and the electroweak correction of $e^-_R e^+_L$ is $12\%$. 
At a CM energy of $1000$ GeV, the electroweak correction of $e^-_L e^+_R$  is $4.0\%$, wheres the electroweak correction of $e^-_R e^+_L$ is $26\%$. 
The $e^-_R e^+_L$ polarization has larger radiative corrections than does the $e^-_L e^+_R$ one. 
Together with the larger cross sections, one can expect smaller systematic errors for the cross-section measurement with the polarized beam than in the non-polarized case.
We note that the full electroweak correction reported here includes a trivial photonic correction from the initial-state photon radiation (ISR)\@.
It is known that the ISR correction can be factorized and be improved using higher-order re-summation\cite{doi:10.1143/PTPS.100.1}. 
Although further improvements of precise predictions of cross sections are possible by means of the parton shower method, we do not use that in this study.
The polarization asymmetry of electroweak corrections may be induced by diagrams involving $W$ bosons\cite{Kou}, i.e., the diagrams shown in Fig.~\ref{Feynman diagrams}. 
In this report, we do not discuss the origin of the radiative-correction asymmetry in detail. 

\subsection{Angular distributions}
\begin{figure}[t]
 	\begin{center}
 		\includegraphics[width={\linewidth}]{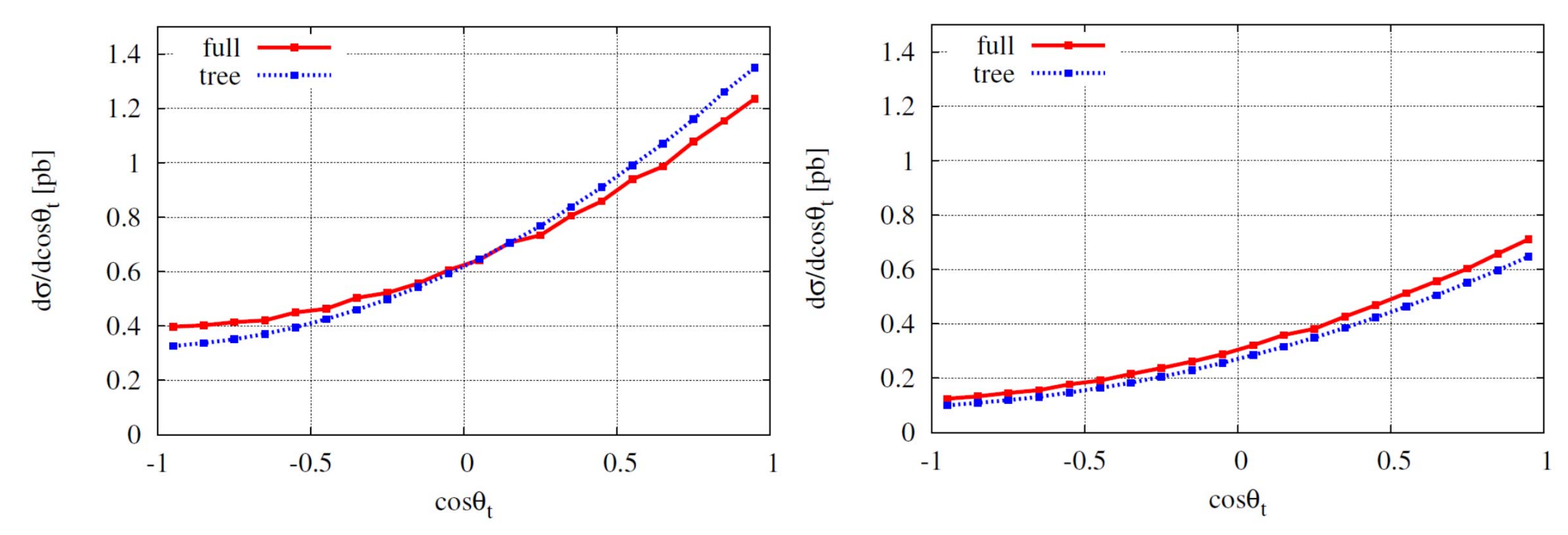}
 		\caption{Angular distributions of the production angle of top quark $\theta_{top}$ at a CM energy of $500$ GeV with $e^-_L e^+_R$ polarization (left) and $e^-_R e^+_L$ polarization (right). 
The dotted lines show tree-level results wheres the solid lines show full electroweak-corrected results.}
 		\label{angular dis LR}
 	\end{center}
 \end{figure}

The angular distribution of the top-pair production has a large forward peak, thus it has a sizable forward-backward asymmetry that allows us to make a good test of the standard theory.
However, radiative corrections may distort the angular distribution as well as the total cross sections.
Angular distributions of the top-pair production with and without radiative corrections at the CM energy of $500$ GeV are shown in Fig.~\ref{angular dis LR} for both $e^-_L e^+_R$ (left figure) and $e^-_Re^+_L$ (right figure) polarizations.
The ISR corrections generally flatten the forward peak because of a smearing effect of the CM system.
One can see this smearing effect clearly in the $e^-_L e^+_R$ polarization case.
Even though the total correction $\delta$ is small at $\sqrt{s}=500$ GeV as mentioned above, the electroweak correction modifies the angular distribution.
A small correction to the total cross section is caused by an accidental cancellation between negative corrections for the forward region and a positive contribution in the backward region.
In contrast, the electroweak correction for the $e^-_R e^+_L$ polarization gives positive corrections in the whole angular region, as shown in the right-hand panel in Fig.~\ref{angular dis LR}.
In conclusion, the observed value of the forward-backward asymmetry is largely affected by the electroweak radiative corrections.
Moreover, the effect of the radiative corrections depends on the spin-polarization of the initial beams.
Therefore, careful investigations of the forward-backward asymmetry are required.

A definition of the forward-backward asymmetry is given as follows.
The forward and back-ward cross sections are defined as $\sigma_F=
 \int_{0}^{1}
 {d\sigma}
/ {d\cos{\theta_t}}
~d\cos{\theta_t}$
 and 
 $\sigma_B=
 \int_{-1}^{~0}
 {d\sigma}
 /{d\cos{\theta_t}}
~d\cos{\theta_t}$, respectively.
Thus, the forward-backward asymmetry is defined by
 $A_{FB}=
( {\sigma_F-\sigma_B})/
( {\sigma_F+\sigma_B})$.
The tree and electroweak corrected values of the forward-backward asymmetry are summarized in Table \ref{4.1}. 
For $e^-_L e^+_R$ ($e^-_R e^+_L$) polarization, the forward-backward asymmetry at tree level is $0.385$ ($0.467$), which becomes $0.317$ ($0.443$) after the full electroweak correction. 
When the design values of polarizations are assumed, the forward-backward asymmetry is determined mainly by the contribution from the $e^-_L e^+_R$ component, as shown in the last row of Table \ref{4.1}.

 \begin{table}[b]
 	\begin{center}
 		\begin{tabular}{|c||r|r|}
 			\hline 
 			$e^-e^+\rightarrow t\bar{t}$ & $A_{FB}$(Tree) & $A_{FB}$(Full)\\
 			\hline
 			$e^-e^+$ & $0.410$ & $0.359$\\
 			$e^-_Le^+_R$ & 0.385 & 0.317\\
 			$e^-_Re^+_L$ & 0.467 & 0.443\\
 			$e^-_L(80\%)e^+_R(30\%)$ & 0.388 & 0.321\\
 			\hline
 		\end{tabular}
 		\caption{Estimated values of the forward-backward asymmetry}\label{4.1}
 	\end{center}
 \end{table}

\subsection{Top-quark decay}
\begin{figure}[t]
 	\begin{center}
 		\includegraphics[width={10cm}]{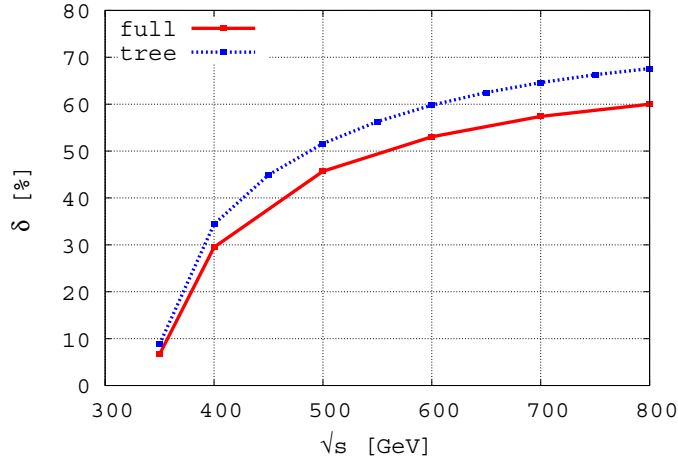}
 		\caption{Top-quark polarization as a function of  CM energy from 300 GeV to 800 GeV for the process $e^-_L e^+_R \rightarrow t \bar{t}$. }
 		\label{top polarization}
 	\end{center}
 \end{figure}

According to the beam polarization, the produced top quarks are also polarized.
The polarization degree is defined as  $\delta_{pol}=({\sigma_L - \sigma_R})/({\sigma_L + \sigma_R})$, where $\sigma_L$ and $\sigma_R$ are the cross sections for creating left-handed and right-handed top quarks, respectively.
The polarization degree depends on the CM energy, as shown in Fig.~\ref{top polarization}.
At tree level, the polarization degree increases from $8.8\%$ at $350$ GeV to $67.6\%$ at $800$ GeV.
At a CM energy of 350 GeV (close to the production threshold), the produced top quark moves slowly and thus its helicity state is easily flipped.
In contrast, at higher energies, the particle moves much faster and the helicity is stable.
That causes the difference in polarization to increase with energy, as shown in Fig.~\ref{top polarization}. 
The full electroweak corrections reduce the polarization degree by roughly $10\%$ in the high-energy region.
The top quark immediately decays into a bottom (b) quark and a fermion pair.
Because the angular and energy distributions of the decay products depend strongly on the top polarization, an exact treatment of the top polarization is necessary.
We discuss the top decay of $t \longrightarrow b\mu^+\nu_{\mu}$ at a CM energy of 500 GeV as a benchmark process.
Because  b-quark tagging is required to identify the top quark experimentally, precise calculation of b-quark distributions  is important.

The number of Feynman diagrams for the six-body final-state $e^- e^+ \rightarrow b \bar{b}\mu^-\mu^+\nu\bar{\nu}$ is too large, thus a full electroweak correction is impossible using the current computing power. 
Instead, we have used a \lq\lq{}narrow width approximation\rq\rq{} for the top-quark production and decay, including the spin correlation exactly. 
The branching ratio of the $b\mu^+\nu_{\mu}$ decay is obtained with the $\cal{O}$$(\alpha)$ correction as follows:
the top width at tree level is calculated to be $\Gamma^{Tree}=1.38$ GeV\@. 
The full electroweak-corrected width is calculated by summing all possible decay channels of $t\rightarrow b l \nu_l$ and $t\rightarrow b q \bar{q}$ as $\Gamma^{Loop}=1.77$ GeV\@.
The partial width of the decay channel to $b\mu^+\nu_{\mu}$ is $0.188$ GeV, thus the branching ratio of this channel is obtained as $10.6\%$ after the $\cal{O}$$(\alpha)$ correction.

\begin{figure}[t]
 	\begin{center}
 		\includegraphics[width={\linewidth}]{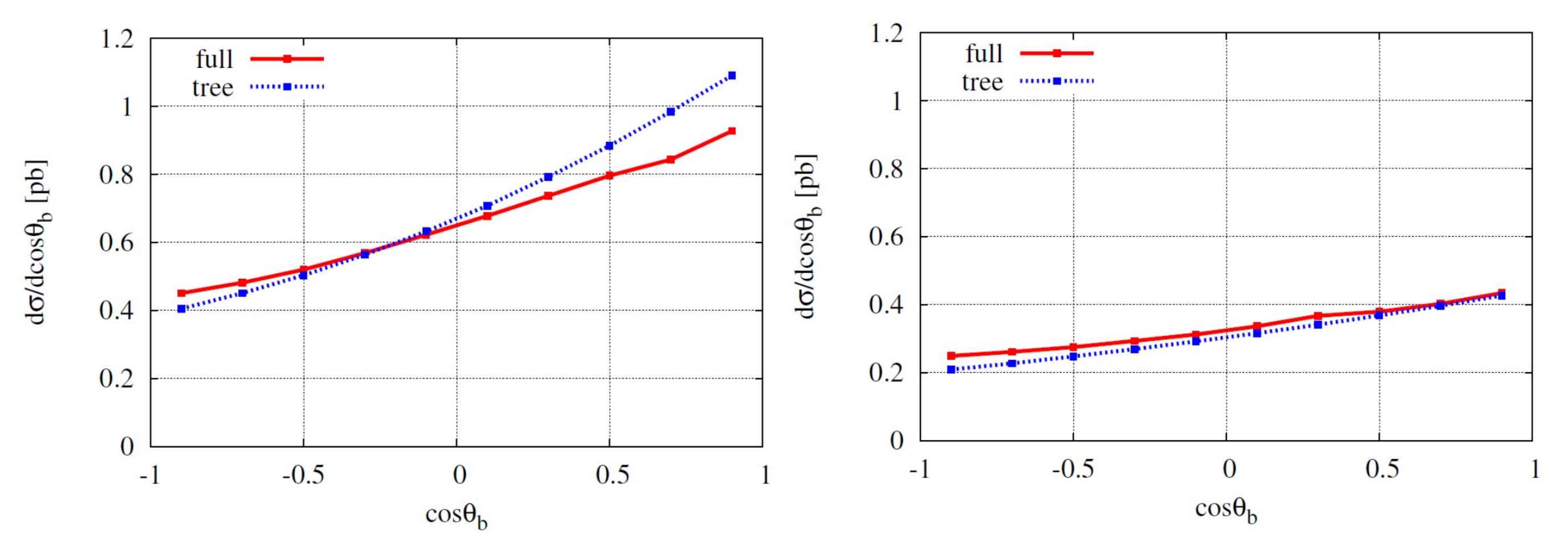}
 		\caption{Angular distributions of b quarks with $e^-_L e^+_R$ (left) and $e^-_Re^+_L$ (right) polarizations. Dotted lines and solid lines show tree and electroweak corrected distributions., respectively.}
 		\label{cosbquarkLR}
 	\end{center}
 \end{figure}

\begin{figure}[t]
 	\begin{center}
 		\includegraphics[width={\linewidth}]{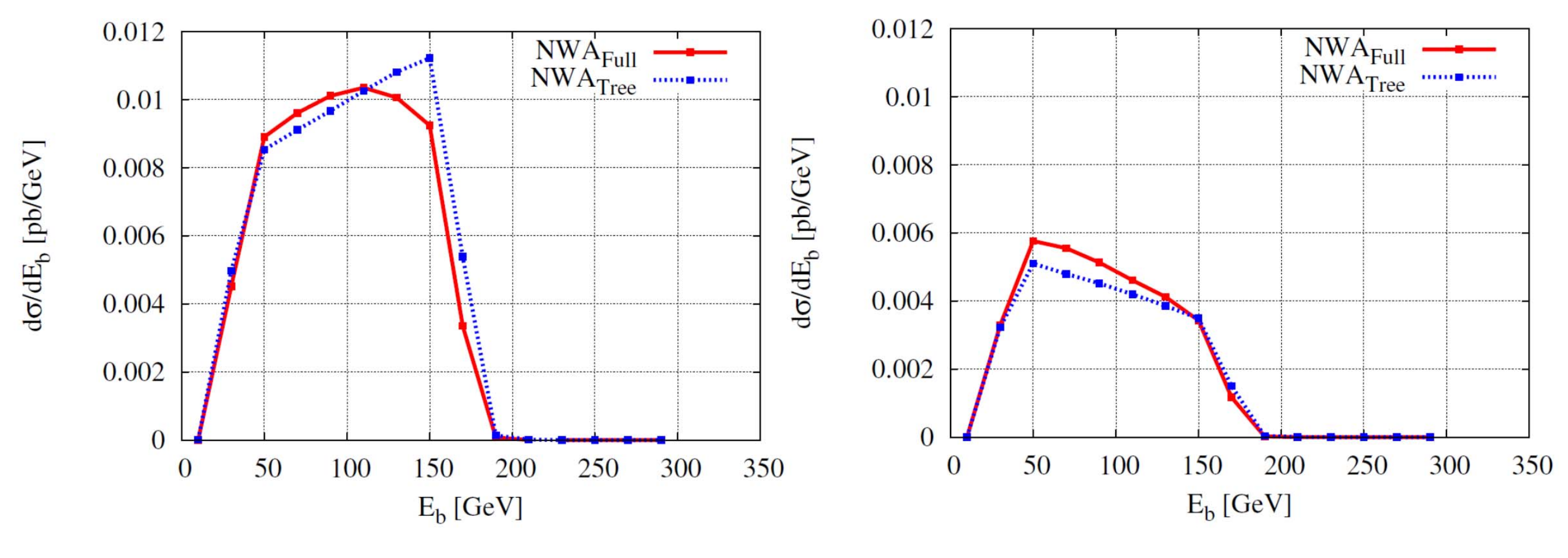}
 		\caption{Energy distributions of b quarks with $e^-_L e^+_R$ (left) and $e^-_Re^+_L$ (right) polarizations. Dotted lines and solid lines show tree and electroweak corrected distributions, respectively.}
 		\label{bquark_energy_LR}
 	\end{center}
 \end{figure}

The total cross section of $N$-body production including a narrow fermion-resonance with mass $m$ and width $\Gamma$, which decays into $N$ bodies, can be expressed as
\begin{eqnarray*}
 	\sigma
 	&=& \frac{1}{flux} \int ~ \abs{\mathcal{M}}^2 d\Omega_N\\
 	&=& \frac{1}{flux} \int 
 	\frac{\abs{\sum_\lambda \mathcal{M}_p u_{\lambda}(q)\bar{u}_{\lambda}(q)\mathcal{M}_d}^2}
 	{(q^2-m^2)^2+m^2\Gamma^2}
 	\frac{dq^2}{2\pi}
 	d\cos{\theta_q} d\varphi_q d\Omega_n d\Omega_{N-n},
\end{eqnarray*}
where $u_{\lambda}$ is the spinor, $q_{\mu}$ is the momentum (off-shell), and $\lambda$ is the spin of the resonance particle. 
The term $d\Omega_n$ denotes an $n$-body phase space, and $\mathcal{M}_p$ and $\mathcal{M}_d$ are the product and decay amplitudes, respectively.
Using an on-shell approximation as $q^2\sim q_0^2=m^2$ for the numerator, the amplitudes can be approximated as
$\tilde{\mathcal{M}}_p^{\lambda}=\mathcal{M}_p u_{\lambda}(q_0)$ and 
$\tilde{\mathcal{M}}_d^{\lambda}=\mathcal{M}_d u_{\lambda}(q_0)$.
Therefore, the total cross section becomes 
\begin{eqnarray*}
 	\sigma 
 	&\simeq &  \frac{1}{flux} 
 	\sum_{\lambda} \int \abs{\tilde{\mathcal{M}}_p^{\lambda}}^2 d\cos{\theta_q} d\varphi_q d\Omega_{N-n}
 	\int \abs{\tilde{\mathcal{M}}_d^{\lambda}}^2 d\Omega_n 
	\int 
 	\frac{1}{(q^2-m^2)^2+m^2\Gamma^2}
 	\frac{dq^2}{2\pi}.
\end{eqnarray*}
We note that the spin correlation is maintained between production and decay.
Integration can be performed over the resonance masses, namely
\begin{eqnarray*}
 	\int \abs{\tilde{\mathcal{M}}_d^{\lambda}}^2 d\Omega_n 
	\int_{-\infty}^{+\infty}~ 
 	\frac{1}{(q^2-m^2)^2+m^2\Gamma^2}
 	\frac{dq^2}{2\pi}
 	&=& \frac{1}{\Gamma} \frac{1}{2m} \int \abs{\tilde{\mathcal{M}}_d^{\lambda}}^2 d\Omega_n,
\end{eqnarray*}
which gives the branching ratio of a specific decay channel.
In reality, calculations are performed using the exact six-body phase space.
The validity of the narrow-width approximation is verified by comparing b-quark distributions obtained by the narrow-width /vise and the exact six-body calculations at tree level.

the angular and energy-distributions of b quarks are shown in Fig.~\ref{cosbquarkLR} and \ref{bquark_energy_LR}, respectively.
For the $e^-_L e^+_R$ polarization case, the decayed b quarks tend to be produced in the forward direction of the top-quark momentum, and in the forward direction for the $e^-_R e^+_L$ polarization.
The angular distributions of the b quarks at tree-level reflect this tendency.
The electroweak corrections distort the angular distribution rather largely in the $e^-_L e^+_R$ polarization case, as shown in the left-had panel of Fig.~\ref{cosbquarkLR}.
In the top-quark rest frame, the b-quark energy is monochromatic (while ignoring the $W$-boson width).
Thus, the energy distribution of the b quarks are a reflection of their angular distribution with respect to the top-quark momentum, after the Lorentz boost due to finite top-momentum.
From this point on view, the energy distribution of b-quarks can be understood intuitively.
Again, the electroweak corrections distort the distribution largely for the $e^-_L e^+_R$ case, as shown in Fig.~\ref{bquark_energy_LR}.

\subsection{QCD correction}
\begin{figure}[t]
 	\begin{center}
 		\includegraphics[width={12cm}]{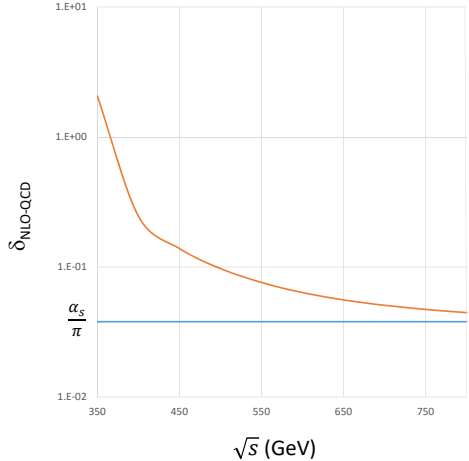}
 		\caption{NLO-QCD correction of the top-pair production process. A strong coupling constant $\alpha_s=0.12$ is used.}
 		\label{QCDcorr}
 	\end{center}
\end{figure}

We have not discussed the QCD correction so far in this report because the QCD correction for the top-pair production  is independent of the beam polarization and simply modifies the total cross section while maintaining the  distributions.
However, the QCD correction is not small at a CM energy of $500$ GeV. 
The formulae used here are summarized in \ref{ap}\@.
While the QCD correction is expected to be $\alpha_s/\pi\simeq3.8\%$ at higher energies it still makes a  contribution of $9.7\%$ to the total cross section at a CM energy of $500$ GeV\@.
While the QCD correction gradually approaches the asymptotic value of $\alpha_s/\pi$ with increase of the CM energy, as shown in Fig.~\ref{QCDcorr}, it still makes a large contribution around a CM energy of $500$ GeV\@.
For future experimental analysis, the QCD corrections around these energies must be investigated more precisely.

%
%
\section{Summary and Conclusions}\label{summary}
In this report, we have present full $\mathcal{O}(\alpha)$ electroweak corrections for the $e^-e^+\rightarrow t \bar{t}$ process associated with the sequential decay $t\rightarrow b \mu\nu_\mu$.
Calculations were performed using the GRACE-Loop system.
The electroweak radiative correction was estimated typically as a level of $10\%$ on the total cross section in the on-shell  scheme for the non-polarized case. 
Wheres the cross section with $e^-_L e^+_R$ polarization was roughly twice that with $e^-_R e^+_L$ polarization at tree level, the radiative correction of the former was smaller than that of the latter.
The electroweak correction with the design polarizations  ($e^-_L=80\%$ and $e^+_R=30\%$) was estimated to be less than $5\%.$
Even though the electroweak correction of the total cross sections was rather small for $e^-_L e^+_R$ polarization, the radiative corrections modified the angular distribution of the produced top quarks. 
The radiative corrections decreased the forward-backward asymmetry of the top-quark production from $0.388$ to $0.321$ for the design polarization.
We also studied the properties of top-quark decay $t \rightarrow b\mu^+\nu_{\mu}$ including the spin correlation.
Both production and decay processes were calculated with $\cal{O}$$(\alpha)$ corrections and combined with using the narrow-width approximation.
We observed the energy distribution of b-quarks to be largely distorted because of the radiative correction.
Therefore, an event generator including radiative corrections for both production and decay with the spin correlation will be necessary for precise measurements in future ILC experiments.
Because the NLO-QCD correction is still large at CM energies $500$ GeV, a precise QCD correction is also desired.

\vskip 0.5cm
The authors wish to thank Prof.~J. Vermaseren for his continuous encouragement and fruitful discussions.
T.U. is supported by the ERC Advanced Grant No.320651 ``HEPGAME'.'
\appendix
\section{QCD correction}\label{ap}
The detailed formulae of the NLO-QCD correction for massive quark-pair production by electroweak interaction are summarized in this Appendix\@.
In following calculations, the standard $\overline{\mathrm{MS}}$ renormalization scheme is used.
After renormalization, a space-time dimension other than four is reinterpreted to regulate the infrared divergence as $d=4-2\varepsilon_{UV}\rightarrow4+2\varepsilon_{IR}$ with $\varepsilon_{IR}>0$.
The NLO-QCD correction consists of three parts: vertex, self-energy, and real-gluon-emission corrections.
The contributions of each part is given separately below.\\

\noindent
{\large \bf Vertex correction}\\
The total vertex correction is given as
\begin{eqnarray*}
\Gamma&=&C_F\left(I_k+I_0\right),
\end{eqnarray*}
where $C_{F}=4/3$ is a color factor.
Each integration term is given as
\begin{eqnarray*}
I_k&=&\frac{\alpha_s}{4\pi}\left\{\frac{-1}{\varepsilon_{IR}}+\left(
L\rq{}+\log{\mu_t}+1+\tilde{\mu}_t\log{\left(-\frac{1-\tilde{\mu}_t}{1+\tilde{\mu}_t}\right)}
\right)
\right\},\\
I_0&=&\frac{\alpha_s}{4\pi}\left\{
\frac{1}{\varepsilon_{IR}}\frac{2\left(2\mu_t+1\right)}{\tilde{\mu}_t}
\log{\left(-\frac{1-\tilde{\mu}_t}{1+\tilde{\mu}_t}\right)}
-1-\frac{2\left(2\mu_t+1\right)}{\tilde{\mu}_t}
\left(
Sp\left(\frac{1}{2}-\frac{1}{2\tilde{\mu}_t}\right)-
Sp\left(\frac{1}{2}+\frac{1}{2\tilde{\mu}_t}\right)
\right)\right.\\ 
&~&+\frac{2}{\tilde{\mu}_t}\log{\left(-\frac{1-\tilde{\mu}_t}{1+\tilde{\mu}_t}\right)}\left(
-2\left(6\mu_t+1\right)\right.\\ &~&
+\left.\left.\left(2\mu_t+1\right)
\left[
L\rq{}+\frac{1}{2}\log{\left(-\frac{1-\tilde{\mu}_t}{1+\tilde{\mu}_t}\right)}+
\log{\left(-\frac{\tilde{\mu}_t(\tilde{\mu}_t+1)}{2}\right)}
\right]
\right)
\right\},\\
&~&L\rq{}=\log{\left(\frac{-s}{\mu_F}\right)},~~\mu_t=-\frac{m_t^2}{s},~~\tilde{\mu}_t=\sqrt{4\mu_t+1},
\end{eqnarray*}
where $\mu_F$ is the factorization energy scale,  $m_t$ is the top-quark mass, and $s$ is the momentum square of a $t\bar{t}$-system.\\

\noindent
{\large\bf Self-energy correction}\\
The self-energy correction appears because of the renormalization scheme.
The top mass that appears here must be interpreted as the $\overline{\mathrm{MS}}$ mass:
\begin{eqnarray*}
\Sigma\left(p^2=m_t^2\right)&=&C_F\frac{\alpha_s}{4\pi}\left\{
\frac{-1}{\varepsilon_{IR}}+\left(L_t-4\right)
\right\},
\end{eqnarray*}
where $L_t=\log{\left({m_t^2}/{\mu_F^2}\right)}$.\\

\noindent
{\large\bf Real-emission correction}\\
The real-emission correction is further separated into two parts: soft-gluon emission and hard-gluon emission.
A threshold energy $k_c$ is introduced to separate soft  and hard emissions. 
The soft-emission corrections are given as
\begin{eqnarray*}
R_{ii}&=&C_F\frac{\alpha_s}{2\pi}\left\{
\frac{1}{\varepsilon_{IR}}-L_k-\frac{1}{\tilde{\mu}_t}
\log{\left(-\frac{1-\tilde{\mu}_t}{1+\tilde{\mu}_t}\right)}
\right\},\\
R_{ij}&=&C_F\frac{\alpha_s}{2\pi}\left\{
\frac{-1}{\varepsilon_{IR}}\left(\frac{2\mu_t+1}{\tilde{\mu}_t}\right)
\log{\left(-\frac{1-\tilde{\mu}_t}{1+\tilde{\mu}_t}\right)}\right.\\
&~&-\frac{2\mu_t+1}{\tilde{\mu}_t}\left.\left(
L_k\log{\left(-\frac{1-\tilde{\mu}_t}{1+\tilde{\mu}_t}\right)}
+
Sp\left(\frac{2}{1+1/\tilde{\mu}_t}\right)-
Sp\left(\frac{2}{1-1/\tilde{\mu}_t}\right)
\right)
\right\},
\end{eqnarray*}
where $L_k=2\log{(2k_c/\mu_F)}$. 
These formulae are obtained via an approximation in which the gluon energy is much smaller that $m_t$.
The hard-emission cross section can be calculated using the GRACE system based on the exact matrix element.
We confirmed numerically that real-emission corrections are independent of $k_c$, whose values are below $1$ GeV.\\

\noindent
{\large\bf Total correction}\\
The NLO-QCD cross section $\sigma_{NLO}$ can be obtained as
\begin{eqnarray*}
\sigma_{NLO}&=&\left\{1+2\left(
R_{ii}+R_{ij}+\mathrm{Re}\left[\Gamma+\Sigma\right]
\right)\right\}\sigma_0+\sigma_g,
\end{eqnarray*}
where $\sigma_0$ and $\sigma_g$ are the Born and hard-emission cross sections, respectively. 
After summing up all contributions, the infrared divergence and $\mu_F$ dependence disappear completely.

%
\bibliography{ref}

\end{document}